\documentclass[conference]{IEEEtran}				
\usepackage[multi-part-units=repeat,detect-all=true]{siunitx}
\usepackage{graphicx}
\usepackage{booktabs}
\usepackage{caption}
\usepackage{listings}					
\usepackage[utf8]{inputenc}
\usepackage{amsmath}
\usepackage{url}
\usepackage{longtable}
\usepackage{lscape}
\usepackage{enumitem}
\usepackage{balance}
\usepackage{fancyvrb}

\renewcommand{\thetable}{\Roman{table}}

\begin{document}

\title{Biting the CHERI bullet:  Blockers, Enablers and Security Implications of CHERI in Defence}

\author{\IEEEauthorblockN{Shamal Faily}
\IEEEauthorblockA{\textit{Defence Science \& Technology Laboratory} \\
Portsdown West, UK \\
sfaily@dstl.gov.uk}
}

\maketitle

\begin{abstract}
There is growing interest in securing the hardware foundations software stacks build upon.  However, before making any investment decision, software and hardware supply chain stakeholders require evidence from realistic, multiple long-term studies of adoption.  We present results from a 12 month evaluation of one such secure hardware solution, CHERI, where 15 teams from industry and academia ported software relevant to Defence to Arm's experimental Morello board.  We identified six types of blocker inhibiting adoption:  dependencies, a knowledge premium, missing utilities, performance, platform instability, and technical debt.  We also identified three types of enabler: tool assistance, improved quality, and trivial code porting.  Finally, we identified five types of potential vulnerability that CHERI could, if not appropriately configured, expand a system's attack surface: state leaks, memory leaks, use after free vulnerabilities, unsafe defaults, and tool chain instability.  Future work should remove potentially insecure defaults from CHERI tooling, and develop a CHERI body of knowledge to further adoption.  This paper was originally presented at the NATO Science and Technology Organization Symposium (ICMCIS) organized by the Information Systems Technology (IST) Scientific and Technical Committee, IST-209-RSY – the ICMCIS, held in Oeiras, Portugal, 13-14 May 2025.
\end{abstract}

\begin{IEEEkeywords}
CHERI, Morello, C, C++, Defence, grounded theory, secure hardware
\end{IEEEkeywords}

\section{Introduction}
\label{sect:intro}

\subsection{Background}

The foundations of computer hardware and software were not designed with modern security requirements in mind.  As Cyber Security concerns have grown, so too has the need to protect the infrastructure society relies on, given the impact that eroded trust could have on the digital economy.  To address these foundations, recent government initiatives have targeted the hardware foundations upon which software stacks are based.  On one hand, doing so is a natural progression to security requirements; hardware might be costly, but investing in it could offset the cost of software security, and the need for continual software updates \cite{tpsh22}.  On the other hand, the hardware supply chain is no less immune to attack than the software supply chain, with a range of well-known invasive, non-invasive, and side channel attack vectors \cite{lev20}.

A secure hardware initiative recently promoted by both the UK and US governments \cite{cisa23,wh24} is CHERI (Capability Hardware Enhanced RISC Instructions): a hardware-software co-design initiative, which enables fine-grained memory protection.  Given claims that 70\% of software vulnerabilities can be attributed to memory safety issues \cite{msrc19}, wide-spread adoption of CHERI could significantly increase cyber resilience.  However, government promotion does not automatically lead to widespread adoption.  For this, several stakeholder interests need to be attended to.  Software supply chain vendors want to understand what the return on investment might be for moving to a new software-hardware solution, based on the challenges and opportunities perceived by software engineers.  Hardware supply chain vendors want to understand the market their products.  These interests  deadlock as evidence to one group of stakeholders cannot be easily obtained without evidence from the other.  Experiments and small case studies can provide some of this evidence, but breaking this deadlock requires evidence from multiple, long-term, convincing evaluations of the technology.

To contribute to this body of evidence, this paper shares the results from a 12-month evaluation of CHERI by 15 teams from industry and academia. Each team was provided with a Morello development board.  This board was produced by Arm as the first industrial quality demonstrator of CHERI \cite{morello}.  The teams were required to port a self-selected body of software to the new hardware.  The ported software was either used in, or applicable to the Defence sector.  The Defence sector may not appear to offer findings generalisable to the wider community, but it does have three characteristics both interesting from a software engineering perspective, and relevant from a software-hardware readiness perspective.  First, many expectations associated with enterprise information technology do not hold in Defence.  Software is often developed for bespoke, embedded, systems, and written in low level, memory unsafe languages like C and C++.  Second, many systems are high-integrity, with both security and safety implications.  Failure could harm not just those direct stakeholders operating the software, but many indirect stakeholders affected by its failure.  As such, software certification is a key concern.  Third, ``legacy-by-design'' software is dominant.  It is not uncommon for software, and the hardware it runs on, to be in operation for decades.  As such, there are large bodies of code that, for operational reasons, cannot be easily ported to other, memory-safe languages. This makes the idea of increased assurance by doing little more than rebuilding software for new hardware an attractive proposition.

The structure of this paper is as follows.  In Section \ref{S:relatedwork}, we briefly introduce CHERI and its hardware and software stack, before presenting the approach taken in Section \ref{S:approach} to run the multi-team evaluation.  In Section \ref{S:results}, we present the results of our analysis.  The implications of these results and the risks introduced by CHERI are considered in Section \ref{S:discussion}.  Finally, in Section \ref{S:conclusion}, we propose two directions for future work to advance wider adoption of CHERI.

\section{Related Work}
\label{S:relatedwork}

\subsection{CHERI}

CHERI is a hardware-software co-design initiative, started in 2010 by the University of Cambridge and SRI International.  CHERI consists of a conventional processor Instruction-Set Architecture (ISA) extended with architectural capabilities to enable fine-grained memory protection and highly scalable software compartmentalisation \cite{wmsn19}.  Compartmentalisation allows data to be protected not just at rest and in transit, but also while it is in use, and without the use of coarser-grained processes or containers.

CHERI provides two forms of memory safety.  \emph{Referential safety} is provided through the protecting the integrity and provenance of pointers against corruption and misuse. \emph{Spatial safety} is provided by preventing a pointer to one current in-memory object being used to access another.  CHERI does not provide \emph{temporal safety}, which would prevent a pointer to one current in-memory object from accessing past or future objects using the same storage (i.e. through a pointer now accessing another context).  Although previous work has considered how this might be supported through capability revocation \cite{figu20}, there are no current plans to incorporate temporal safety into CHERI due to the complexity it would introduce \cite{wdb20}. 

CHERI is not the first attempt to address memory safety in hardware.  For example, Arm introduced a Memory Tagging Extension (MTE) to the Armv8.5-A processor \cite{gret18}.  The extension assigned tags to each memory allocation, which would need to accompany each memory access via a pointer; feedback can be provided to the operating system in the event of memory errors.  However, what differentiates CHERI from previous work in software and hardware memory safety is its non-reliance on secrets or probabilistic techniques, a focus on minimising disruption to large, pre-existing bodies of existing code, and a willingness to depend on modest changes to hardware architecture and microarchitectures to enable requisite security properties \cite{wccd24}.

\subsection{CHERI hardware}

Early CHERI cores extended the Microprocessor without Interlocked Pipeline Stages (MIPS) architecture, and targeted field-programmable gate arrays (FPGAs) \cite{wccd24}.  This was the initial hardware target for developing the CHERI approach.

The Morello development board was produced by Arm as the first industrial quality demonstrator of CHERI.  Morello is an experimental board, and only a limited number have been produced.  The Morello multicore system on chip (SoC) incorporates four CHERI-enabled Neoverse-N1 processors \cite{gbwm23}.  The board implements the Morello Arm architecture extension \cite{morello_arm}; this adds instructions for features like setting pointer bounds, and sealing / unsealing capabilities.

More recently, Microsoft have developed the CHERIoT (Capability Hardware Extension to RISC-V for Internet of Things) ISA \cite{amch23}, which was designed for smaller, embedded targets.  The CHERIoT-Ibex RISC-V microcontroller \cite{cibex} implements the CHERIoT ISA, and has formed the basis of the Sonata board \cite{sonata}.  However, because this study preceded the release of Sonata, discussion of this board is not considered further in this paper.

Arm's development of Morello and lowRISC's development of Sonata was funded through the UK Digital Secure by Design (DSbD) programme \cite{dsbd}: a five-year programme funded in 2019 by the UK Industrial Strategy Challenge Fund (ISCF).  In addition to funding technical development of CHERI, the programme has also run periodic \emph{All Hands} meetings to help grow the emerging CHERI ecosystem.

\subsection{CHERI software}

Although there has been recent work porting Rust \cite{haco23} to CHERI, the bulk of software created for CHERI is written in C and C++.  CHERI C and C++ \cite{wdb20} are variants of their respective languages, but the presence of capabilities make these more than just ``memory safe C and C++''.  The open-source Clang/LLVM compiler and LLD linker \cite{cllvm} was extended to support the generation of \emph{pure capability} machine code, where pointers are implemented as CHERI capabilities.  These are unforgeable tokens of authority, and implemented with twice the width of native integer pointer types, and incorporate elements for validity, memory bounds, permissions data, and an indicator of whether or not the capability can be modified or dereferenced.  To support environments where non-CHERI aware machine code may be present, CHERI also supports the \emph{hybrid} C/C++ where only selected pointers are implemented as capabilities \cite{wdb20}.

To date, experimental support for CHERI has mainly been incorporated into two operating systems.  The first is CheriBSD: a capability enabled extension of the FreeBSD operating system \cite{cheribsd}.  CheriBSD is comparatively mature, with several thousand memory-safe packages already ported to it, including a full KDE-based desktop stack.  The second is Morello Linux \cite{mlin}, which implements a pure capability kernel-user Application Binary Interface (ABI), supported by C and C++ libraries (e.g. musl libc) and tooling (e.g. Morello LLVM).  

\section{Approach}
\label{S:approach}

To gather evidence for adoption, identify novel usage ideas, and support further related research \& development, Dstl acquired a number of Morello units to enable industrial/academic evaluation of CHERI.  Dstl ran a themed competition  \cite{dasa} for teams to access Morello boards and project funding for up to 12 months to address one or more of three challenge area: (i) code porting, (ii) software compartmentalisation, (iii) innovation.  A total of 29 proposals were submitted by teams from industry and academia, 16 were recommended for funding, and 15 (D1 - D15) proceeded to contract.  

Throughout the 12 month period, each team was supported by a Dstl technical partner.  The technical partner was responsible for independently confirming the fitness for purpose of the work carried out by each team, and that the outputs produced met their own processes for technical quality and review.  The technical partner met periodically online, and occasionally in-person, with each team to review progress, discuss experiences and challenges, and comment on draft technical deliverables.

Because of the range, maturity, and size of software corpora being targeted, together with differing expectations around performance and maintainability, we decided that quantitative metrics would unsuitable as measures of adoption.  Therefore, in addition to delivering technical artifacts, each team was required to submit a final report on their experiences with CHERI/Morello.  These reports were analysed using Grounded Theory \cite{cost08}.  Grounded Theory is a well-established  methodology for making sense of and conceptualising qualitative data.  Validity of theory emerging from a Grounded Theory analysis is obtained through transparency of the approach taken, and the soundness of the arguments constituting the theory.

To capture a broad range of experiences, each team was asked to structure their reports to encompass the following areas and questions:

\begin{itemize}
	\item Aims: what security did teams hope to achieve, and what Validation \& Verification approach was taken?
	\item Setup: what experience with Morello did teams have, and how was information about CHERI/Morello obtained?
	\item Development: for the targets chosen by teams, what challenges were faced with dependencies, the available documentation, tooling, and porting?  Additionally, how was security evaluated, what performance trade-offs were observed, and what benefits did CHERI/Morello bring?
	\item Reflections: how much did CHERI/Morello meet team expectations, and what features could have the most and least impact on security?
\end{itemize}

Coding of the reports was carried out by the author.  The author has significant and recent C++ software development experience, significant Cyber Security and Software Engineering experience, experience porting software to CHERI and Morello within a Defence project, and an internationally recognised, publication track record applying Grounded Theory to Cyber Security contexts, e.g. \cite{fafl10,famc15}.  Following open coding of an initial sample of 10 reports (D1-D10), 204 codes (i.e. themes) based on 237 quotations were obtained.  Following de-duplication, axial coding, and selective coding, and the inclusion of an additional report, 75 codes based on 268 quotations had been identified.  Initial results from this model were shared with technical partners, and two additional subject matter experts, following which the remaining reports were analysed.  Some additional codes were identified after analysis of the 11th report (D11), and theoretical saturation, i.e. the point that no new codes were identified, was obtained after analysis of the 12th report (D12).  After a final round of selective coding to support the model write-up, the final model constituted of 88 codes and 321 quotations.

Due to commercial restrictions around the release of content from the team reports, no direct quotations are included in the paper.  However, to back up key points from the analysis, paraphrased examples are provided.

\section{Results}
\label{S:results}

\subsection{Supplier background and aims}
\label{S:supplier_background}

\begin{table}[]
\centering
\begin{tabular}{@{}lll@{}}
\toprule
\textbf{Team} & \textbf{Sector} &  \textbf{Experience}\\ \midrule
D1  & Civil Security & CHERI awareness \\
D2  & Industrial Control & Unix development \\
D3  & Defence & TAP alumni  \\
D4  & System Integration & Wide-ranged  \\ 
D5  & System Integration & Unix development \\
D6  & Maritime & Unix development \\
D7  & Defence & Unix development \\
D8  & Automotive  & TAP alumni  \\
D9  & Defence  & Unix development      \\
D10 & System Integration & TAP alumni  \\
D11 & Higher Education & TAP alumni \\
D12 & System Integration & TAP alumni \\
D13 & Higher Education & TAP alumni \\
D14  & Higher Education & Unix development \\
D15  & Aerospace & CHERI awareness\\ \midrule
\end{tabular}
\caption{Sectors and pre-existing CHERI experience of teams}
\end{table}

As Table I illustrates, pre-existing knowledge of CHERI within teams varied, but could be characterised as one of four of categories of expertise:

\begin{itemize}
\item CHERI awareness: Awareness of CHERI obtained through material provided by the DSbD programme.  In some cases, such teams also had limited experience with Unix and C.
\item Unix development:  Experience developing software for one or more flavours of Unix/Linux.  In some cases, teams also had some low-level experience developing and maintaining drivers, with some knowledge of Assembly.
\item TAP alumni  : DSbD Technology Access Programme (TAP) \cite{tap} alumni, with current or recent experience working with the Morello board.
\item Wide-ranged: Extensive experience developing software for CHERI.
\end{itemize}

Only a single team had \emph{Wide-ranged} expertise, and a small number of teams were limited to \emph{CHERI awareness}.  The remainder of teams met one of the remaining two categories.  

\begin{table}[]
\resizebox{\columnwidth}{!}{%
\begin{tabular}{@{}llll@{}}
\toprule
\textbf{Team} & \textbf{Product} &  \textbf{Languages} & \textbf{Target OS} \\ \midrule
D1  & Smart camera & C, C++ & CheriBSD \\
D2  & UAV infrastructure libraries & C, C++ & CheriBSD \\
D3  & Vehicular monitoring system & C & CheriBSD  \\
D4  & Web stack & C, C++ & CheriBSD  \\ 
D5  & IoT framework component & C & CheriBSD \\
D6  & Autonomous system components & C++ & Linux \\
D7  & General component libraries & C & CheriBSD \\
D8  & Vehicular middleware libraries  & C & CheriBSD \\
D9  & Cross Domain Solution  & C, C++, Assembly & CheriBSD  \\
D10 & Wireless access component & C, C++ & CheriBSD  \\
D11 & RTOS and Rust compiler extensions & C, Rust, Assembly & Bare metal \\
D12 & OS kernel & C, C++, Assembly & Bare metal \\
D13 & Python interpreter & C & Bare metal \\
D14  & UAV middleware libraries & C, C++ & CheriBSD \\
D15  & UAV flight control components & C++ & CheriBSD \\ \midrule
\end{tabular}}
\caption{Products ported by teams, programming languages used, and target operating system}
\end{table}

Table II summarises the body of software ported by each team.  These fell into one of three categories.  The first were stand-alone applications; some of these run on well-established desktop or server operating systems, but were targeted to a host supporting CHERI, e.g. CheriBSD.  Other stand-alone applications include an Operating System kernel and Real-Time Operating System (RTOS), which run on ``bare-metal'' hardware.  The second are components of some other application.  These were other software libraries linked into the larger application, or stand-alone executables coupled to a larger framework, i.e. via API or network protocols.  The last category was a ``stack'' consisting of multiple applications or components at different levels, e.g. the software stack necessary to run web applications, and software applications using middleware.

No team explicitly set out to make their products secure.  They wanted to port their respective applications or components to CHERI/Morello.  This was, however, done with the expectation that using capabilities would lead to improved security outcomes.  Despite these expectations, teams largely relied on custom functionality tests, unit tests, or some combination of both.  There was little specific consideration of security in the Validation \& Verification approaches taken beyond some ad-hoc out-of-bound memory testing.

\subsection{Blockers}

\emph{Blockers} are the inhibitors found by the teams, which contributed to slow progress and, in many cases, a reduction of the pre-agreed project scope.  Six classes of blocker were identified during this analysis.

\begin{itemize}
\item Dependencies: people and technology upon which teams' work explicitly or implicitly depends. 
\item Knowledge premium: the scarcity, i.e. premium, of knowledge about CHERI, and the technology necessary to develop software for it.
\item Missing utilities: technology teams rely on for software development that was either unavailable or available in a degraded or unstable form.  
\item Performance impact:  the impact of CHERI/Morello on the performance of ported software.
\item Platform instability: instability of CHERI/Morello and its host operating system during the building and running of ported software.
\item Technical debt: the effort necessary to adapt software for the target platform and its host operating system.
\end{itemize}

\subsubsection{Dependencies}
\label{S:dep}

Teams found two classes of significant dependency: social and technical ecosystems, and tool chain instability.  The first component of ecosystem dependency was a dependency on the open-source ecosystem.  In some cases, the teams' choice of target meant that, during porting, there were no significant open-source dependencies because the target was well isolated, or the team had an intimate knowledge of the dependencies.  However, some teams faced significant issues because open-source dependencies they assumed would be present were not.  For example, D2 encountered problems installing a particular package in Python using \verb_pip_ due to dependencies such as \verb_numpy_ and \verb_scipy_.  However, these dependencies had not been ported to the CheriBSD operating system used by the team.

The second component was a dependency on other communities.  One of these is the DSbD community, but others are communities associated with commercial and open-source dependencies.  Many such communities are international and, even if they were aware of the DSbD community and sympathetic of its goals, their interests are orthogonal.  For example, D6 attempted to port an open source middleware product to CHERI, but the focus of that open source community was not CHERI but an overhaul of the product that consumed discussion across multiple websites and Discourse channels. 

Suppliers were also dependent on the stability of the tools they were using.  Suppliers were often using tools they had some familiarity with, but with an unexpected level of instability.  For example, D11 encountered frequent crashes or unusual behaviours with the CHERI community's Eclipse IDE.  Software development often entailed writing and editing assembly code to get to a state where, while not correct, the debugger and Eclipse no longer crashed.

\subsubsection{Knowledge premium}
\label{S:kp}

Almost every team found their progress inhibited by one or more of five documentation blockers:

\begin{itemize}
\item Outdated documentation: guides or scripts providing non-working instructions. In some cases, this was due to search engines indexing out-of-date content.  
\item Compartmentalisation gaps: teams were unable to find adequate material on how to apply software compartmentalisation.
\item Documentation distribution: teams were left to forage for information they needed on sites hosted by different organisations. The published information was not always consistent, particularly because each site made assumptions about the host Operating System, i.e. CheriBSD in documentation hosted on one site, and Linux in documentation hosted on another.
\item Missing examples: teams missed non-trivial examples of how to use capabilities, which would be relevant to their projects.  This blocker was particularly problematic for those using compartmentalisation, or relying on the use of Assembly or C for interfacing hardware.
\item Build gaps:  teams found little help troubleshooting build problems, or guidance on cross-compilation.  In some cases, teams were reliant on examples found in GitHub repositories or archived posts on mailing lists.
\end{itemize}  

As indicated in Section \ref{S:supplier_background}, the pre-existing knowledge teams brought to their projects played a role in the knowledge premium they had to pay.  In the case of outdated documentation, the fast moving nature of dependent open source projects such as CheriBSD also played a part. A significant factor is the expectation made by authors of documentation about those consuming it, i.e. their expertise was the same as theirs.  For example, instructions provided with the Morello referenced the need to build firmware, without indicating what firmware and how.  The expectations were particularly opinionated when teams were left to work with the same code repositories being actively maintained.  For example, D8 encountered difficulties building software when changing the Git repository branch they were working on.  On obtaining support, D8 was directed to not change the branch but instead build using the Morello branch instead.  Guidance on what branches to use and their stability were not documented.  

Problems with documentation, particular its distributed nature, led to shared confusion in the community about different aspects of CHERI.  This was especially the case for compartmentalisation.  For example, one DSbD All Hands meeting featured an open session on compartmentalisation; this was attended by Dstl, and several teams.  At the session, teams perceived a lack of uniform definition for what compartmentalisation was, and several attendees described having trouble understanding the documentation associated with it.  

\subsubsection{Missing Utilities}

Suppliers were explicitly asked to comment on which tools they missed while porting their software.  Their responses feel into one of three categories of blocker.  

First, teams working with CheriBSD were unable to use Integrated Development Environments (IDEs) they were accustomed to.  Some teams were left working directly with unfamiliar makefiles and debuggers directly from the terminal.  Others had to find ways of sharing directories across machines, so they could use IDEs on Linux machines.  

Second, although debuggers were available for CHERI/Morello, some teams found them less functional than expected when working with CHERI specific features.  For example, D9 noted that crashes within compartments were not associated with source code within the debugger.  Pointer addresses of crashes had to be manually adjusted by the base offset of the compartment, and then searched for in object dumps to find the affected function.  As such, the use of \verb_printf_ became the only viable debugging method, and this sometimes relied on relaying to an implementation outside of the compartment.

Finally, many teams missed expected tools from their usual build tool chain.  These included code analysers, linters, and performance analysis tools tailored for CHERI/Morello.  In some cases, a stable C++ development environment was also expected but missing.  This led to some teams resorting to cross-compilation for CHERI/Morello on an off-target development environment. 

\subsubsection{Technical Debt}
\label{S:td}

Four types of technical debt blocker were encountered.

Given the changes to pointer size, challenging assumptions made about pointers within code bases was expected by teams.  In some cases, this entailed changing pointer types to ensure portability, but - in some cases - assumptions about pointer sizes were used in build tools to make decisions about a target's architecture.  For example, D6 found that CMake examined the pointer size on target systems when determining if a system was 32-bit or 64-bit.  As such, because CHERI is a 64-bit platform with 128-bit pointers, this broke CMake detection.

Because of the introduction of a new signal (SIGPROT) for handling capability exceptions, teams were required to implement defensive code and new exception handling mechanisms.  This was necessary not just for code susceptible to buffer overflows, but behaviour previously considered acceptable that CHERI now considered unsafe.  If not addressed, such behaviour might be dangerous to security and safety.  For example, D7 noted the possibility that an unexpected operating condition could hit the boundary of a sandbox and trigger an exception.  They considered this a design error that should have been found and fixed during design and verification, but careful consideration of exception handling would be needed to ensure that the system recovered in a safe and secure way.

A claimed benefit of CHERI is its application to legacy code with minimal changes.  While the changes made by teams were comparatively minor, some refactoring of legacy code took place - either as an opportunity to improve its quality, or because changes were forced due to incompatibility resulting from the use of LLVM.  For example, D6 found that one of their target's dependencies would not build out-of-the-box with the Morello CMake tool chain due to references to \verb_ltstdc++fd_: a legacy C++ library for backwards compatibility with the \verb_filesystem_ library on older compilers.

Finally, the use of a different operating system led some teams to make architectural changes to their software to account for now absent software dependencies.  For example, D5 found that a significant task of their porting exercise was to entirely decouple their application's logic, so it could be driven by a platform independent API.  This included UDP stack handling, serial port access, multi-threading, and thread safe message queues.  This entailed porting software to Linux, determining the software worked as expected, and migrating the code to CheriBSD.

\subsubsection{Performance Impact}
\label{S:perf}

Given the experimental nature of the CHERI/Morello, the performance of the ported software was found to be adequate.  However, notably reduced performance was observed by teams.  Four classes of performance blocker were identified.

\begin{itemize}
\item Algorithmic changes: either due to software architectural changes, or changes to the machine code generated by the tool chain.
\item Inference speed: reduced inference speed for machine learning applications and inefficient query plans in database applications.  
\item Network latency: at times, a reduced throughput of 35\% was noted for one popular open-source RPC framework.
\item Allocation speed: the type and size of memory allocated from the heap had a notable impact on performance.  One team noted a reduction in execution time of around 20\% during the processing of small chunks of memory.
\end{itemize}

The performance issues could not be easily attributed to either the CHERI ABI or the Morello micro-architecture.  Moreover, latency could also have been introduced by bugs during the software porting process, particularly for those applications comparatively high up the software stack.  Optimisation opportunities do, however, exist should the nature of the targeted software be well understood.  For example, on considering the software they were porting, D4 found that some functions did not access any state in their respective libraries, and were either pure functions or operated only on a state object provided as an argument.  As such, those functions could potentially be executed  without a domain transition by incurring minimum or no loss of security guarantees.

\subsubsection{Platform Instability}
\label{S:pi}

Two types of platform instability blocker were identified: CHERI instability and CHERI immaturity.  

The first class, CHERI instability, related to instability associated with CHERI/Morello.  Hardware instability was evident in several ways, including undefined behaviour when accessing registers on the Morello, resetting of the processor, and unpredictable boot times.  CHERI incompatibility arose through ABI incompatibility, and linker issues due to incompatible object types.  Compartmentalisation immaturity resulted from the need to use low-level features to implement isolation; teams expected such features to be available via compiler options or through interfaces similar to containers or hypervisors.

The second class, CheriBSD immaturity, concerned instability associated with the CheriBSD operating system.  Driver instability characterised some known issues that were subsequently patched in the kernel, but not all driver problems were known.  For example, D8 found that WiFi support appeared to be unstable.  There were unpredictable periods of extremely slow response to input, while, at other times, the connection appeared stable and fast, seemingly irrespective of network conditions.

Linux / CheriBSD variations were not limited to the missing dependencies described in Section \ref{S:dep}; they included changes to interfaces, e.g. support for Portable Operating System Interface (POSIX) mqueues and different types for file descriptors.  And while CheriBSD was initially assumed to be a stable, capability aware operating system, teams identified multiple instances of OS instability.  These ranged from minor issues like missing man pages and unavailable 3rd party packages, to more serious issues where CheriBSD ports of drivers caused kernel panics. These resulted from unhandled CHERI exceptions caused by pre-existing bugs in FreeBSD.  Finally, build troubleshooting issues included troubleshooting of CMake errors due to Free/CheriBSD incompatibility, as well as random segmentation faults in LLVM while running the CheriBSD cheribuild.py script.  In some cases, teams adopted the practice of porting code to Linux and mainstream processors as a staging activity before porting to CheriBSD and Morello.

\subsection{Enablers}

\emph{Enablers} are features of CHERI that teams felt added value to their work.  Three types of enabler were identified.

First, teams found that the CHERI-enabled tools such as compilers and debuggers were notably more effective at detecting memory problems than other tools, particularly for issues that might otherwise be difficult to debug.  As suggested in Section \ref{S:kp}, a contributing factor for this might be the expectations set by the tool authors of what might be useful diagnostic information for other developers like them.  Were such helpful diagnostic information not identifiable at compile time, CHERI exceptions could follow at runtime.  While the developers of CHERI tools would know this, application developers new to this technology might not.

Second, the quality of the code generally improved.  This could be partly attributable to the payment of technical debt as indicated in Section \ref{S:td} or porting software to a platform less tolerant of build-time errors.  Another factor was the role of CHERI as a tool for finding memory bugs at compile time, rather than runtime or after exploitation of the software.  For example, D5 believed there was a strong argument that production code intended for non-CHERI enabled hardware could benefit from running in test mode on the Morello platform.  The benefit of the Morello and an exhaustive test harness would be to build confidence that issues related to capability exceptions could be identified and fixed before entering service in production code, i.e. even if the final code did not run on CHERI enabled hardware, testing on CHERI hardware would help identify problems.

Finally, the exercise of porting software was comparatively trivial, given the sizes of the teams' respective code bases.  For example, the size of a library ported by D6 was  216650 Lines of Code (LOC) after being configured and built by the CHERI-specific tool chain.  An additional ten lines of code were needed to create the tool chain, to make the library compile for CHERI.  A further 39 lines were changed to allow the library, and the library examples, to be built for CHERI.  However, as suggested in Section \ref{S:td}, some assumptions about pointers could be difficult to unpack, and the time taken to identify affected code was occasionally lengthy.  As such, measuring ease of migration based on lines changed also does not reflect the time taken to make the changes.  Consequently, the exercise of porting code at lower-levels of the software stack might be comparatively less trivial.

\subsection{Security implications}
\label{S:sec}

CHERI was designed to potentially remove the memory attack surface of software.  However, novel technology can introduce complexity with the potential to increase attack surface.  As indicated in Section \ref{S:td}, some of this complexity results from the need for extra exception handling, but idioms might also arise when specific interaction with CHERI primitives is required.  Some of this complexity also results from CHERI usage patterns inferred from the documentation or examples, particularly for compartmentalisation.  For example, D3 resorted to using dynamic loading instead of dynamic linking for a library they ported.  This design pattern was inherited from some example code from a Software Development Kit (SDK), where the documentation indicated that the choice was made to provide greater flexibility and portability across platforms.  It was not, however, obvious to D3 what the benefits of this approach were.

Five classes of increased attack surface, which constitute potential vulnerabilities to CHERI, were identified:

\begin{itemize}
\item State leak: leakage of state information into library compartments through shared object data, particularly through the use of C++ classes across compartments.
\item Memory leak: memory leaks resulting from inappropriate memory management of compartmentalisation primitives.
\item Use after free: reuse of a pointer to a previously deallocated object \cite{cwe416}, but one for which a valid capability remains.
\item Unsafe defaults: default compiler flags provide lower levels of assurance to maintain compatibility.
\item Tool chain instability: as indicated in Section \ref{S:dep}, unexpected instability could lead to insecure code changes to ensure applications remain running.  For example, one team encountered a link-time problem but was unsure about whether it was a bug that could be worked around, or an intentional security feature.
\end{itemize}

Given its effectiveness at providing spatial safety, such vulnerabilities could contribute to a false sense of security.  Decision makers may not appreciate that CHERI is unable to prevent attacks due to flawed logic and improper access control \cite{cwe284}.  Moreover, while CHERI guards against attacks on memory misuse, it does so at the cost of potential denials of service.  This could constitute a safety hazard depending on the role of the software being protected.

\subsection{Competing Technology}

Teams perceived three competing approaches for achieving similar levels of protection to CHERI:  

\begin{itemize}
\item Memory safe languages: languages such as Rust claim to offer the performance advantages of C with improved memory handling.  CHERI does, however, facilitate hardware guarantees for \verb_unsafe_ code blocks, although developers are starting to focus on the security of such blocks \cite{hkwa23}. 
\item Software memory protection: many operating systems support techniques like Address Space Layout Randomization (ASLR), and compilers such as \verb_gcc_ support memory sanitisation to detect out-of-bounds and use after free vulnerabilities.
\item Hypervisors: hypervisors implement compartmentalisation at a higher level of abstraction than CHERI but with greater ease of use.  However, like memory safe languages, hypervisors could be supplemented by CHERI to obtain stronger levels of assurance.  The use of certain compartmentalisation patterns, such as library-based compartmentalisation or compartmentalisation based on policy, could lead to precise control of software boundaries without unduly sacrificing developer user experience.
\end{itemize}

The performance impact described in Section  \ref{S:perf} fed this perception.  So too did the potential non-applicability of compartmentalisation, e.g. in applications where privilege separation cannot be usefully applied, and where memory attacks are not a significant feature of threat models.  Market adoption uncertainty also played a role in perceptions about competing technology.  

Each approach assumes a software and hardware footprint similar to Morello. However, many teams expressed an interest in using CHERI on embedded targets that lack the resources of conventional hardware, and may not support the current generation of memory safe language and hypervisors.  For example, D8 plans to continue monitoring CHERI technology.  They are keen to evaluate CHERIoT-based FPGAs with a view to converting to low level OSes, and attempting to add temporal memory safety at the hardware level; this was considered more achievable on hardware smaller than Morello. 

What appeared not to influence team perceptions about competing technology were the challenges of migrating an existing system, e.g. re-writing software in a memory safe language, or re-architecting an application to use container-based compartmentalisation.

\section{Discussion}
\label{S:discussion}

\subsection{Contestable compatibility and security claims}
\label{S:contestable}
A claimed key benefit of CHERI is its application to legacy C or C++ programs with minimal changes \cite{dsbd_how}.  Most teams achieved some level of success porting their respective code bases, and the team effort employed may have been less than porting their respective code bases to a memory safe language.  Nonetheless, the technical debt associated with porting software to CHERI described in Section \ref{S:td} suggests this key benefit claim is contestable.

Claims that the majority of operating system vulnerabilities are due to memory safety issues that CHERI could address are also contestable.  CHERI appears only to mitigate 2 of the current Top 10 CWEs \cite{cwe_top10} by default.  While the top CWE (Use after Free) \emph{could} be mitigated by CHERI based on recently improved capability revocation performance \cite{figu24}, doing so would likely entail breaking compatibility - a key benefit of CHERI.  As indicated by Section \ref{S:td}, this could potentially lead to additional technical debt that broadly increases the attack surface.

\subsection{Documentation and tooling as reverse salients}

As an experimental platform, CHERI/Morello is not fully formed, so instability and immaturity is to be expected.  However, as an industry demonstrator, it is also reasonable to examine how ready for adoption the different components of this technology are.  

Seminal work in technology innovation \cite{hugh83} found that innovative system growth relies on correcting \emph{reverse salients}: imbalances that occur when some parts of a system develop faster than others.  The results from Section \ref{S:results} suggest that, while the CHERI architecture and compiler technology continues to advance at pace, documentation and supplemental tooling that supports software engineering remains left behind.  Failing to address these reverse salients could have two negative security implications.  

First, delays delivering documentation and tooling of acceptable quality shifts an additional security burden to developers, who - as suggested in Section \ref{S:contestable} - may already be paying indirect costs for adopting CHERI.  The burden documentation and tooling causes has also been corroborated by work commissioned by Discribe (the Digital Security by Design Social Science Hub+) \cite{alhs23}.  In such circumstances, Herley \cite{herl09} indicates that it would then be rational for developers to seek knowledge from sources where the cost-benefit pay-off is in their favour.  This could push developers to potentially untrusted sources of knowledge like Stack Overflow \cite{stackoverflow} for advice and code samples of unknown provenance; such advice and samples could harbour vulnerabilities \cite{fbxs17}.

Second, if the technology readiness of CHERI continues to advance at the cost of the reverse salients, pre-existing challenges with knowledge asymmetries, where knowledge and ignorance is spread unequally across the software development ecosystem \cite{pifz17}, could become malignant.  Asymmetric information about CHERI could grow a market for professional services, but it could also grow a security ``market for lemons'' \cite{mol} where prospective consumers are unable to effectively evaluate security knowledge about CHERI \cite{anmo06}.  This could lead to the market failures that government initiatives hoped to address.

\subsection{Risks to Software Certification}

Software certification is a barrier for technology exploitation in Defence.  Engineering for CHERI effectively, particularly for compartmentalising software, without increasing a system's certification surface, appears to be a complex system integration problem.  For example, CHERI has open-source dependencies where certification evidence, such as requirements, architectural models, and Validation \& Verification plans, are absent.  The lack of this evidence is prohibitive to standards such as the Software Considerations in Airborne Systems and Equipment Certification (DO-178C) \cite{do178c}, and restrict the ability of assessors to assure soundness of the development process \cite{tud21}.  Moreover, some of the platform stability issues described in Section \ref{S:pi} such as unhandled CHERI exceptions and unexplained board behaviour make both hardware and software certification problematic.

\section{Conclusion}
\label{S:conclusion}

In this paper, we presented the results from a 12 month evaluation of CHERI by 15 independent teams working with Defence applicable software.  As such, this work constitutes the first-long term study of CHERI readiness.

The role of this paper is not to propose recommendations based on these results; there is no certainty of successful adoption across industry even if all the blockers were addressed, and all enablers exploited.  Moreover, as indicated in Section \ref{S:discussion}, adopting CHERI does not come without risk.  We do, however, suggest two future directions for reducing these risks, and advancing the adoption of CHERI, particularly for Defence contexts.  

First, claims that CHERI can address the majority of security vulnerabilities at minimal cost fails to acknowledge a key tension between security and maintainability. To address this tension, permissive but potentially insecure defaults should be removed, so memory protection is ``by default and by design''.  While this will increase the cost of adoption, these costs may be comparatively small given the other costs incurred.  The costs may also be acceptable for Defence systems, particularly those running on embedded platforms.  This change will also expose problems and opportunities that may otherwise be missed when prospective users are shielded from the consequences of full memory protection.

Second, to further shed light on CHERI knowledge asymmetries, a range of education and training material is needed on the productive design, development, and maintenance of CHERI-enabled software.  This material should also be explicit about the pre-existing knowledge expected of those wishing the join the CHERI community.  Doing so will not only address pre-existing knowledge asymmetries, and potentially identify missing knowledge gaps, it will make an agreed CHERI body of knowledge as open and accessible as its source code.  Such a body of knowledge should be accessible not only to professional developers, but also for undergraduate and postgraduate students.  This ensures that CHERI foundations are embedded into the education of the next generation of software and security engineers.  Such a body of knowledge will also aid software certification by indicating the knowledge needed by suitably qualified and experience personnel required to build, operate, and maintain CHERI-enabled systems.

\section*{Acknowledgements}

Although this paper has a single author, two other members of Dstl made substantive contributions.  However, due to the sensitivity of their work, they have asked not to be publicly disclosed as authors.  This document is an overview of UK MOD sponsored research. The contents of this document should not be interpreted as representing the views of the UK MOD, nor should it be assumed that they reflect any current or future UK MOD policy.

\textcopyright~ Crown copyright (2024), Dstl. This information is licensed under the Open Government Licence v3.0. To view this licence, visit \url{https://www.nationalarchives.gov.uk/doc/open-government-licence/version/3}. Where we have identified any third party copyright information you will need to obtain permission from the copyright holders concerned. Any enquiries regarding this publication should be sent to: Dstl.

\balance
\bibliographystyle{IEEEtran}
\bibliography{mrefs}

\end{document}